\documentclass[epj]{svjour}
\usepackage{epsfig} \usepackage{psfig} \usepackage{times}

\begin{document}
\title{Probing the interactions of charmed mesons with nuclei in
 $\bar p$ induced  reactions \footnote{supported by DFG, RFFI and
Forschungzentrum J\"{u}lich.}}
\author{ Ye.S. Golubeva\inst{1},
W. Cassing\inst{2}, L.A. Kondratyuk\inst{3}} \institute{Institute
for Nuclear Research, 60th October Anniversary Prospect 7A, \\
117312 Moscow, Russia \and Institut f\"ur Theoretische Physik,
Universit\"at Giessen, \\ D-35392 Giessen, Germany   \and
Institute of Theoretical and Experimental Physics, B.\
Cheremushkinskaya 25,
\\ 117259 Moscow, Russia}
\date{Received: date / Revised version: date}

\abstract{ We study the perspectives of resonant and nonresonant
charmed meson production in $\bar{p} + A$ reactions within the
Multiple Scattering Monte Carlo (MSMC) approach. We calculate the
production of the resonances $\Psi(3770), \Psi(4040)$ and
$\Psi(4160)$ on various nuclei, their propagation and decay to $D,
\bar{D}, D^*, \bar{D}^*, D_s, \bar{D}_s$ in the medium and vacuum,
respectively. The modifications of the open charm vector mesons in
the nuclear medium are found to be rather moderate or even small
such that dilepton spectroscopy will require an invariant mass resolution of a few MeV.
Furthermore, the elastic and inelastic interactions
of the open charm mesons in the medium are taken into account, which can be related to
$(u,d)$-, $s$- or $c$-quark exchange with nucleons. It
is found that by studying the $D/\bar{D}$ ratio for low momenta in the laboratory
($\leq 2-2.2$ GeV/c) as a function of target mass $A$ stringent constraints on the
$c$-quark exchange cross section can be obtained. On the other hand, the ratios
 $D^-_s/D^+_s$ as well as $D/D^-_s$
and $D/D^+_s$ at low momenta as a function of $A$ will permit to fix independently
the strength of the $s$-quark exchange reaction in $D^-_s N$ scattering. }


\authorrunning{Ye. S. Golubeva et al.} \titlerunning{Interactions of
charmed mesons with nucleons in the $\bar p A$ reaction}

\maketitle

\section{Introduction}
 In the last decade  the dynamics of charm quark degrees
of freedom have gained sizeable interest especially in the context
of a phase transition to the quark-gluon plasma (QGP) where
charmed meson states should no longer be formed due to color
screening \cite{QM96,QM97,QM99,Satz,Satz2}. However, the
suppression of the $c\bar{c}$ vector mesons in the high density
phase of nucleus-nucleus collisions might also be attributed to
inelastic comover scattering (cf.
\cite{Cass95,Cass99,Vogt99,Seattle98} and Refs. therein) if the
corresponding charmonium-hadron cross sections are in the order of
a few $mb$ \cite{Haglin,Ko}.  Also the $J/\Psi N$ cross section is
not known sufficiently well \cite{Bernd} since the photoproduction
data suggest a value of 3--4 $mb$ \cite{Huefner}, while the
charmonium absorption on nucleons (at high relative momentum) in
$p+A$ and $A+A$ reactions is conventionally fitted by 6--7~$mb$
\cite{Seattle98,Dimo97} which might relate to the cross section of
a pre-resonance charmonium state with nucleons before the actual
hadronic state is formed. Present estimates for the $J/\Psi$
formation times range from 0.3 -- 0.5 fm/c \cite{Huefner}, but
additional data from $\bar{p}$ induced reactions on nuclei, where
charmonium can be formed on resonance with moderate momenta
relative to the target nucleons, should help to pin down the
present uncertainties \cite{Seth}. Such studies might be performed
experimentally at the HESR, which is presently proposed as a
future facility at GSI \cite{GSIfut}.

In a previous study we have explored the perspectives of charmed
meson - nucleon scattering in the $\bar{p} d$ reaction
\cite{Cass00} where the charmonium is produced on resonance with
the proton in the deuteron and may scatter with the spectator
neutron. Another possibility is also the reaction $\pi^+ d
\rightarrow J/\psi p p$ as discussed by Brodsky and Miller in Ref.
\cite{Brodsky}. Here we investigate $\bar{p} A$ reactions and
study the production  of the resonances $\Psi(3770), \Psi(4040)$
and $\Psi(4160)$ in various nuclei, their propagation and decay to
$D, \bar{D}, D^*, \bar{D}^*, D_s, \bar{D}_s$ in the medium and
vacuum, respectively. Furthermore, the elastic and inelastic
interactions of the open charm mesons in the medium are followed
up in the Multiple Scattering Monte Carlo (MSMC) approach to study
the dominant medium effects as a function of the target mass $A$.

Our work is organized as follows: In Section 2 we will briefly
describe the MSMC approach and evaluate the formation cross
sections for the resonances  $\Psi(3770), \Psi(4040)$ and
$\Psi(4160)$ on protons and nuclei. The fractional decay of these
resonances to open charm mesons in the medium and vacuum,
respectively, is calculated in Section 3 whereas the dynamics of
the open charm mesons is described and studied in Section 4.
Section 5 concludes this paper with a summary and discussion of
open problems.

\section{Resonance production and decay in $\bar{p} p$ and $\bar{p} A$ reactions}
We here examine the possibility to measure the in-medium life time
(or total width) of the resonances $V=(\Psi(3770), \Psi(4040)$
$\Psi(4160)$) produced in $\bar{p} A$ reactions. Before doing so
one has to determine first the resonance properties in vacuum,
i.e. their production in $\bar{p} p$ reactions. To this aim we
describe the vector meson spectral function by a Breit-Wigner
distribution
\begin{equation}
A_V(M^2) = N_V \frac{M
\Gamma_{tot}^V(M)}{(M^2-M_V^2-\Re\Pi_V(M))^2 + M^2
\Gamma_{tot}^2(M)}, \label{BW}
\end{equation}
where $M_V$ denotes the vacuum pole mass, $\Re\Pi_V$ is the vector
meson self energy in the medium - which vanished in the vacuum -
and $N_V \sim 1/\pi$ (for small width $\Gamma_{tot}$) is a
normalization factor that ensures $\int dM^2 A(M^2) =1$. The total
width $\Gamma_{tot}$ is separated into vacuum and medium
contributions as
\begin{equation}
\Gamma_{tot}(M) =  \Gamma_{vac}(M) + \Gamma_{coll}(M),
\label{gamma}
\end{equation}
where \begin{equation} \Gamma_{vac}(M) = \sum_c \Gamma_c(M)
\label{vac}
\end{equation} with $\Gamma_c(M)$ denoting the partial
width to the decay channels $c \equiv D\bar{D}, D\bar{D}^*, D^*
\bar{D}, D^* \bar{D}^*, D_s \bar{D}_s,  D_s^* \bar{D}_s,  D_s
\bar{D}^*_s, D_s^* \bar{D}_s^*$, respectively. The in-medium
collisional width is determined from the imaginary part of the
forward vector-meson nucleon scattering amplitude as
\begin{equation}
\Gamma_{coll} = \frac{4 \pi}{M} \Im f_V(0) \rho_A, \label{coll}
\end{equation}
where $\rho_A$ denotes the nuclear density. Furthermore, $\Im
f_V(0)$ is determined via the optical theorem by the total
vector-meson nucleon cross section $\sigma_{VN}$ that will be
specified below. The real part of the vector meson self energy
$\Re \Pi_V(M)$ in the medium might be calculated by dispersion
relations, however, QCD sum rules for $c \bar{c}$ vector states
suggest that the 'mass shift' $\Re \Pi/(2 M_V)$ at normal nuclear
matter density is only in the order of a few MeV \cite{Klingl} due
to a small coupling of the $c, \bar{c}$ quarks to the nuclear
medium. We thus will assume $\Re \Pi_V$ = 0 throughout the
following study.

\subsection{Vacuum properties}
Whereas the pole masses of the vector mesons $M_V$ are
approximately known e.g. from $e^+e^-$ annihilation \cite{Bai}
this does not hold for the total widths and especially branching
ratios. To this end we have to model the vacuum decays, that enter
the spectral function (\ref{BW}) via (\ref{vac}), using available
information from the PDG \cite{PDG}. In each channel $c$ the
relative decay to mesons $m_1$ and $m_2$ is described by a matrix
element (squared) and phase space, i.e.
\begin{equation}
\Gamma_c \sim |{\cal M}_c|^2 p_c^3, \label{channel}
\end{equation}
where $p_c$ is the meson momentum in the rest frame of the
resonance. In case of the $\Psi(3770)$ only the lowest channel to
$D \bar{D}$ contributes such that the matrix element in
(\ref{channel}) can be fixed by the total width at half maximum
which we take as $\Gamma_{FWHM} =$ 25 MeV \cite{PDG}. For the
$\Psi(4040)$ and $\Psi(4160)$ we adopt $\Gamma_{FWHM} =$ 50 MeV
and $\Gamma_{FWHM} =$ 80 MeV, respectively \cite{PDG}. The
branching ratios for the $\Psi(4040)$ show a strong anomaly
favoring the decays to channels with vector mesons $D^*$. The PDG
\cite{PDG} quotes $\Gamma(D^0 \bar{D}^0)/\Gamma(D^{*0} \bar{D}^0 +
c.c.) = 0.05 \pm 0.03$ and $\Gamma(D^{*0}
\bar{D}^{*0})/\Gamma(D^{*0} \bar{D}^0 + c.c.) = 32 \pm 12$, where
the phase space factor $\sim p_c^3$ already has been divided out.
The origin for such a large variation in the matrix elements is
not known so far. We use the actual numbers (within the range of
uncertainty) $$\frac{{\cal M}^2_{D\bar{D}}}{{\cal M}^2_{D^*
\bar{D}}} = 0.08; \ \frac{{\cal M}^2_{D^*\bar{D}^*}}{{\cal
M}^2_{D^* \bar{D}}} = 22; $$
\begin{equation}
\frac{{\cal M}^2_{D\bar{D}}}{{\cal M}^2_{D_s \bar{D}_s}} = 3; \
\frac{{\cal M}^2_{D_s \bar{D}^*_s}}{{\cal M}^2_{D_s \bar{D}_s}} =
12.5; \ \frac{{\cal M}^2_{D_s^* \bar{D}^*_s}}{{\cal M}^2_{D_s
\bar{D}_s^*}} = 22, \label{4040}
\end{equation}
that provide a reasonable description to the $e^+e^-$ formation
data from Ref. \cite{Bai}. The resulting spectral function $S(M)=2
M A(M)$ for the $\Psi(4040)$ (with a pole mass of 4.08 GeV) is
shown in Fig. 1 (upper part) in terms of the thick solid line as
well as the mass dependent branching ratios for the different
channels $c$ (including the phase-space factor $\sim p_c^3$). It
is clearly seen that for the ratios (\ref{4040}) the $D^*
\bar{D}^*$ decay opens up at $M \approx$ 4.03 GeV and dominates
the spectral function for higher masses. Furthermore, the decays
to $D_s, D_s^*$ are strongly suppressed such that this resonance
does not qualify for the production of $(s\bar{c})$ or $(c
\bar{s})$ mesons.
\begin{figure}[h]
\centerline{\psfig{figure=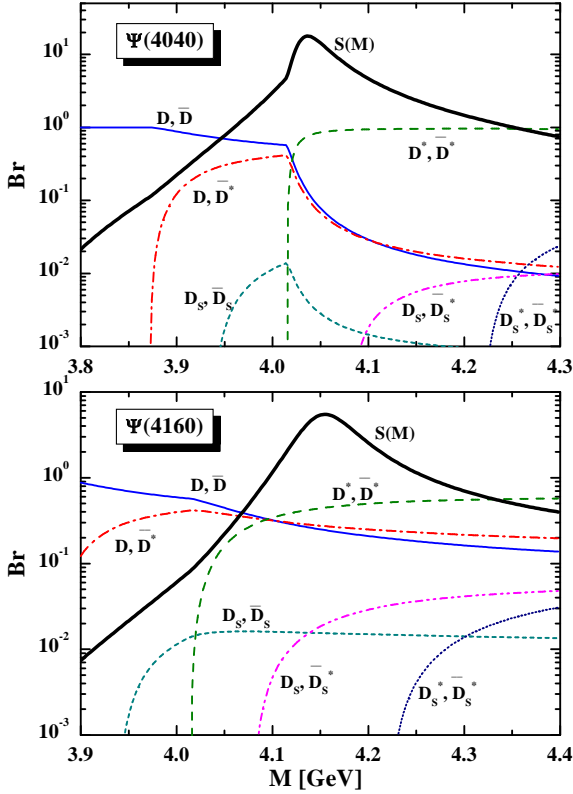,width=8.5cm}}
 \caption{The spectral functions $S(M)$ for the $\Psi(4040)$ and $\Psi(4160)$
(thick lines, not normalized) and the mass differential branching
ratios for both vector mesons within the model described in the
text.}
 \label{bild1}
\end{figure}

Apart from the total width of 80 $\pm 25$ MeV the properties of
the $\Psi(4160)$ are not well known experimentally. For the decay
channels we thus adopt a 'counting rule' for high energy open
charm production that results from PYTHIA calculations (cf. Fig. 1
in  Ref. \cite{Cass01}), i.e. $$ \frac{{\cal M}^2_{D\bar{D}}}{{\cal M}^2_{D^*
\bar{D}}} = \frac{1}{3}; \ \frac{{\cal M}^2_{D^*\bar{D}^*}}{{\cal M}^2_{D^*
\bar{D}}} = 3; \ \frac{{\cal M}^2_{D\bar{D}}}{{\cal M}^2_{D_s \bar{D}_s}} = 3;
$$
\begin{equation}
\frac{{\cal M}^2_{D_s\bar{D}^*_s}}{{\cal M}^2_{D_s \bar{D}_s}} = 3; \
\frac{{\cal M}^2_{D_s^*\bar{D}^*_s}}{{\cal M}^2_{D_s \bar{D}_s^*}} = 3.
\label{4160}
\end{equation}
This is presently nothing but an educated guess (at low energies)
and must be controlled by future data. The resulting spectral
function $S(M)=2 M A(M)$ for the $\Psi(4160)$ is shown in Fig. 1
(lower part) in terms of the thick solid line as well as the mass
dependent branching ratios for the different channels $c$. It is
clearly seen that for the ratios (\ref{4160}) the $D^* \bar{D}^*$
decay still dominates the spectral function, however, the decays
to $D_s, D_s^*$ are no longer so strongly suppressed as in case of
the $\Psi(4040)$.

The production of the vector mesons in $\bar{p} p$ then can be
described by Breit-Wigner resonance formation on the basis of
the spectral function $A(M^2)$ (\ref{BW}) - employing the proper
kinematics - provided that the branching of $\bar{p}p \rightarrow
V$ is known,
\begin{equation}
 \sigma_V(M) = \frac{3}{4} Br_{p\bar{p}
\rightarrow V} 4 M A_V(M) \Gamma^{tot}_V(M) \frac{\pi^2}{k^2},
\label{cross}
\end{equation}
where the factor 3/4 stems from the ratio of spin factors and $k$
denotes the momentum of the $p$ (or $\bar{p}$) in the cms. In
fact, the branching ratio is rather uncertain for the cases of
interest here. Following our previous work \cite{Cass00} we adopt
$Br(p\bar{p} \rightarrow \Psi(3770)) \approx 2 \times 10^{-4}$
which is very similar to the branching of the $\Psi(2S)$ state.
Due to a lack of detailed knowledge we assume this branching ratio
also to hold for the $\Psi(4040)$ and $\Psi(4160)$. The resulting
cross sections for the mesons $V$ in $\bar{p} p$ reactions are
displayed in Fig. 2 (upper part) as a function of the antiproton
kinetic energy $T_{\bar{p}}$ in the laboratory reflecting the
resonance structure (\ref{BW}) and the kinematics from
(\ref{cross}).

\begin{figure}[h]
\centerline{\psfig{figure=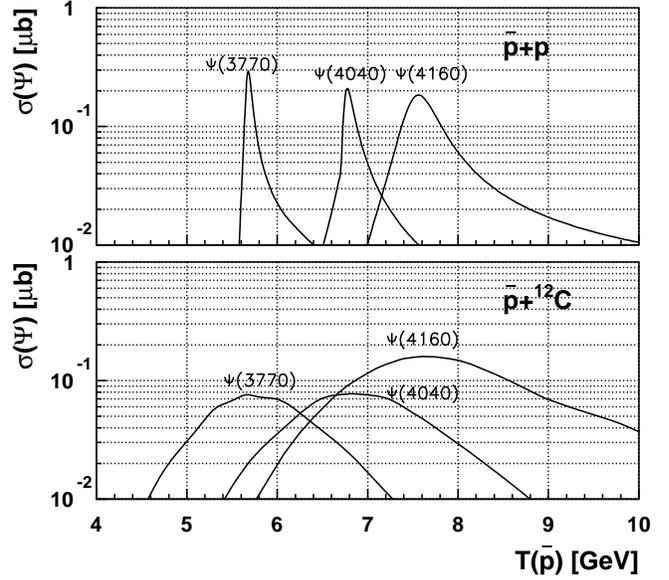,width=8.5cm}}
 \caption{The calculated cross sections for $\Psi(3770)$, $\Psi(4040)$ and
$\Psi(4160)$ for $\bar{p}p$ (upper part) and $\bar{p}+^{12}C$
reactions (lower part) as a function of the antiproton kinetic
energy in the laboratory.}
 \label{bild2}
\end{figure}

\subsection{$\bar{p} A$ reactions}
In order to simulate events for the reaction $\bar{p}A \rightarrow
V $ we use the Multiple Scattering Monte Carlo (MSMC) approach. An
earlier version of this approach -- denoted as Intra-Nuclear
Cascade (INC) model -- has been applied to the analysis of $\eta$
and $\omega$ production in $\bar p A$ and $p A$ interactions in
Refs.~\cite{Golub92,Golub93,Golub96,Golub97}. The production of
the hidden charm vector resonances on nuclei then can be evaluated
within the MSMC by propagating the antiproton in the nucleus and
folding the elementary $\bar{p} p \rightarrow V$ production cross
section with the local momentum distribution at the annihilation
point. We have calculated the latter quantity in the local
Thomas-Fermi approximation.

The numerical results for the formation cross sections of the
resonances $V$ on $^{12}C$ are shown in the lower part of Fig. 2
as a function of $T_{\bar{p}}$ indicating that the resonance
formation is further smeared out by Fermi motion and the maximum
in the cross section becomes reduced inspite of the larger
$\bar{p}$ annihilation cross section on $^{12}C$. The suppression
factor in the maximum of the cross section - per elementary
reaction - is of the order (see Ref. \cite{Farrar})
\begin{equation} S.F. \simeq \frac{\pi \Gamma_{R} m_N}{(k_F M_{R})},
\end{equation} where
$\Gamma_R, M_R$ denote the vacuum width and mass of the produced
meson, whereas $k_F$ is the target Fermi momentum. This leads to
factors of $\sim$ 0.075, 0.14 and 0.21 for $\Psi(3770),
\Psi(4040)$ and $\Psi(4160)$ per $p\bar{p}$ reaction,
respectively. On the other hand, the production on nuclei roughly
scales with the target proton number as $Z^{0.6}$ (see below) such
that the cross section for $^{12}C$ in the maximum for the
$\Psi(4160)$ is roughly the same as in $\bar{p}p$ annihilation,
while it is lower for the narrower resonances $\Psi(3770)$ and
$\Psi(4040)$.

Necessary parameters for a Monte Carlo (MC) simulation of
rescattering are the elastic and inelastic $V N$ scattering cross
sections and slope parameters $b$ for the differential cross
sections $d\sigma_{el}/dt$, which are approximated by
\begin{equation} \frac{d \sigma_{el}}{dt} \sim \exp(bt), \end{equation}
where $t$ is the momentum transfer squared. These parameters as
well as the masses of the rescattered particles determine the
momentum and angular distributions of the particles in the final
state. As in our previous study \cite{Cass00} we use $b$= 1
GeV$^{-2}$ for $D,\bar{D}+N$ and $VN$ scattering as an educated
guess. Furthermore, the inelastic cross sections of the vector
mesons with nucleons have to be specified. Since the relative
momenta in the $V-N$ system for rescattering are in the order of a
few GeV one might use high energy geometric cross sections for the
dissociation to open charm mesons. For all the three resonances
studied here we adopt an inelastic cross section of 11 $mb$ as
well as $\sigma_{el} = 1\ mb$. The inelastic cross section of 11
$mb$ is larger by about a factor of 2 in comparison to the
$J/\Psi, \Psi(2S)$ cross sections (see introduction) due to the
larger size of the higher lying resonances.

As mentioned above, the resonances $V$ are propagated without any
mean-field potential. This approximation should be well fulfilled
since the expected potentials are only in the order of a few MeV
\cite{Klingl} while the energy of vector mesons in the lab. frame
is a couple of GeV. Furthermore, due to the high mass of the
resonances we neglect any intrinsic formation time $\tau_F$ for
the resonances.

The resonances are propagated with their actual mass $M$ -
selected by MC according to the spectral function (\ref{BW}) -
with velocity ${p}_V/E_V$ and decay in time according to the
differential equation
\begin{equation} \frac{d P_V}{dt} = - \frac{1}{\gamma} \Gamma^V_{tot}(M)
P_V(t)
\end{equation}
with the total width (\ref{gamma}), while $\gamma$ denotes the
Lorentz $\gamma$-factor. Their decay to
 $D\bar{D}, D\bar{D}^*, D^*
\bar{D}, D^* \bar{D}^*, D_s \bar{D}_s,$ $  D_s^* \bar{D}_s, $ $
D_s \bar{D}^*_s,$ $ D_s^* \bar{D}_s^*$ is, furthermore, selected
by Monte-Carlo according to the mass differential branching ratios
displayed in Fig. 1.

Due to the rather moderate collision rates involved (for
$\sigma_{VN}^{tot}$ $ \approx \ 12 mb$) the life time of the
vector mesons $\Psi(3770), \Psi(4040)$ and $\Psi(4160)$ is not
reduced sizeably in the nuclear medium relative to the vacuum
(assuming the open charm mesons not to change their properties in
the medium). This reduction then amounts to about 15\% for the
$\Psi(3770),$ to $\sim$ 8\% for the $ \Psi(4040)$ and to $\sim$ 5
\% for the $\Psi(4160)$, respectively. On the other hand, the
$D,\bar{D}$ mesons might change their effective mass in the
nuclear medium as advocated in Refs. \cite{Alex99,Sibirtsev,Haya}.
Especially for dropping $D,\bar{D}$ effective masses in the medium
the decay width of the vector mesons becomes enhanced at finite
density $\rho_A$. This effect is most pronounced for the
$\Psi(3770)$ since its pole mass is close to the $D,\bar{D}$
threshold of $\sim$ 3.73 GeV. A drop of the $D, \bar{D}$ meson
masses both by 50 MeV at density $\rho_0$ would imply an increase
of the decay width by about a factor of 2.5 according to phase
space. Thus in-medium modifications of the open charm mesons might
alter the life time of the $\Psi(3770)$ more effectively than
collisional broadening. However, in-\newline  medium potentials
are most pronounced at low momenta and decrease in size with
increasing momentum. Thus it is very questionable if $D, \bar{D}$
mesons will have sizeable potentials for relative momenta of 3--5
GeV/c with respect to the nucleus at rest. Thus, in the following,
we do not speculate any further on such medium mass modifications
of $D, \bar{D}$ mesons at high relative momentum and continue our
calculations for vacuum properties of the open charm system.

The open charm mesons from the decaying vector mesons have a light
quark $q=(u,d)$ or strange $s$ quark content apart from the
$\bar{c}$ or $c$ quark. In the constituent quark model we get
${D}^+ = (c\bar{d})$, ${D}^- = (\bar{c} {d})$, $\bar{D}^0 =
(\bar{c}{u})$, ${D}^0 = (c\bar{u})$, ${D}^+_s = (c\bar{s})$,
${D}^-_s = (\bar{c}{s})$ and the same composition for the related
vector states. Now light quark exchanges with nucleons ($(uud)$ or
$(udd)$) have different strength as it is well established
experimentally from $K^+ p$ and $K^-p$ reactions. Whereas the $K^+
(u \bar{s})$ scatters only elastically with nucleons at low
momentum - since the $\bar{s}$ cannot be exchanged with a light
quark of a nucleon - the $K^- (s\bar{u})$ cross section is
dominated by resonant $s$-quark transfer leading to strange
baryons such as $\Lambda$ or $\Sigma$ hyperons. Similar relations
are expected to hold for the $D, \bar{D}$ mesons, where especially
the light quark contribution should give much larger cross section
on nucleons than the $c\bar{c}$ vector resonances. This analogy is
based on $SU(4)_{flavor}$ symmetry \cite{Lin} which, however,
might be broken substantially in view of the different geometrical
sizes of $K$- and $D$-mesons. At present this is an open question,
which has to be settled by experiment.

For our calculations  we adopt the following cross sections -
taken as constants in the momentum regime of interest - $$
\sigma^{el}_{{D} N} = \sigma^{el}_{\bar{D} N} = 10 \
 mb; \ $$ $$ \sigma^{inel}_{\bar{D} N} \approx 0; \
\sigma^{inel}_{{D} N} = 10 \ mb;$$ $$ \sigma^{el}_{{D}_s N} =
\sigma^{el}_{{D}_s N} = 3 \ mb; $$
\begin{equation}\sigma^{inel}_{{D}^+_s N} \approx 3 \ mb; \  \sigma^{inel}_{{D}^-_s N}
= 10 \ mb. \label{parameters}
\end{equation}
Here the inelastic cross sections of $D$-mesons refer to $c$-quark
exchange reactions with nucleons. Furthermore, charge ex\-change
reactions like ${D}^+ n \leftrightarrow {D}^0 p$ or ${D}^- p
\leftrightarrow \bar{D}^0 n$ are incorporated with a constant
cross section $\sigma_{q exc.}$ = 12 $mb$. For our exploratory
study we assume the same cross sections for the related open charm
vector mesons $D^*, \bar{D}^*$ etc.

\section{Resonance production in $\bar{p} A$ reactions}
Within the model described in Section 2 we are now in the position
to calculate the cross section of the $c \bar{c}$ resonances in
$\bar{p} A$ reactions for all targets of interest and can obtain
the information about in-medium resonance decays (for densities
$\rho_A \geq 0.03 fm^{-3}$) or vacuum decays, respectively. In
this respect we show in Fig. 3 the calculated cross section for
the $\Psi(3770)$ at $T_{lab}$ = 5.7 GeV for the targets $^{12}C$,
$^{64}Cu$, $^{108}Ag$ and $^{207}Pb$ as a function of target mass
$A$. The solid line represents the power law
\begin{equation}
\sigma_\Psi \sim A^\alpha \label{scal}
\end{equation}
with $\alpha$ = 0.55, which nicely describes the total
$\Psi(3770)$ yield. The coefficient $\alpha$ is less than 2/3 - as
expected from the total $\bar{p}$ annihilation cross section - due
to fact that only reactions with protons can lead to uncharged
resonances and the Fermi smearing is more effective in heavy than
in light nuclei. The fraction of $\Psi(3770)$, that decay to
$D\bar{D}$ inside the nucleus (D ins) is only a few \% in case of
the $^{12}C$ target, however, reaches about 50\% for the $Pb$
target. Thus for a heavy target the in-medium properties of the
$\Psi(3770)$ may well be studied experimentally by dilepton
spectroscopy exploiting the dilepton branching of $7 \times
10^{-4}$ of this resonance. However, as mentioned above, a
collisional broadening of its spectral function by $\sim$ 15 \% is
not expected to provide a clear signal relative to its vacuum
decay unless the experimental mass resolution is in the order of
1-2 MeV.
\begin{figure}[h]
\centerline{\psfig{figure=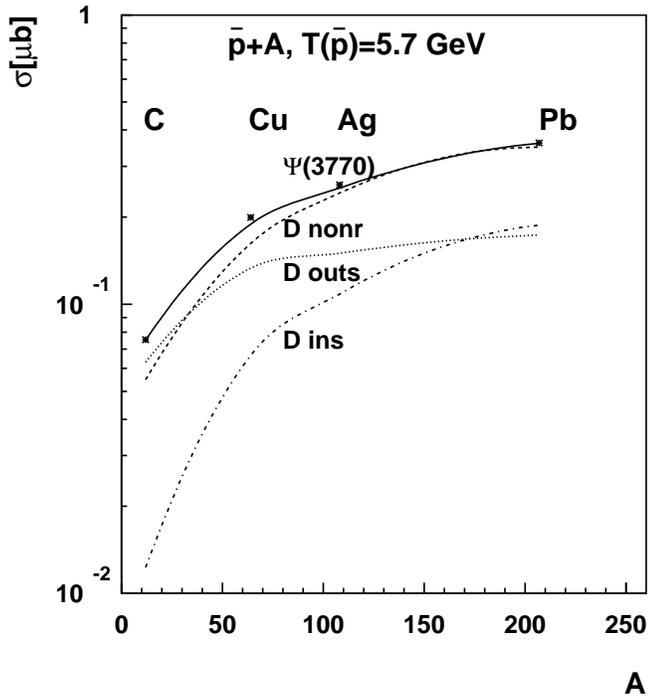,width=8.5cm}}
 \caption{The calculated cross sections for $\Psi(3770)$ at $T_{lab}$ = 5.7 GeV
 as a function of the target mass $A$. The dotted line (D outs) denotes the contribution
 from 'outside' resonance decays whereas the dash-dotted line (D ins) shows the
 resonance contribution from 'inside' decays. The solid line reflects the
 scaling law (\ref{scal}) for $\alpha$ = 0.55 for the total contribution from the resonance decay.
 The dashed line (D nonr) corresponds to the estimated nonresonant
 $D, \bar{D}$ cross section.}
 \label{bild3}
\end{figure}

Apart from the resonant production of open charm mesons they may
also be produced directly as $(c\bar{q})$ and ($\bar{c}q$) pairs
in $\bar{p}N$ annihilation. This channel strongly dominates in
annihilation reactions on neutrons in the nucleus since the
charmonium production is forbidden by charge conservation at low
energy above threshold. At higher invariant energies a pion might
balance the charge in the $\Psi + \pi$ final channel. For our
estimates we have employed the Regge-model analysis of Ref.
\cite{Kaidalov} for the elementary $\bar{p} N \rightarrow D,
\bar{D}$ cross section as also used by Sibirtsev et al. in Refs.
\cite{Alex99,Sibirtsev}. The resulting cross section of
nonresonant open charm production is shown in Fig. 3 as a function
of the target mass $A$ in terms of the dashed line (D nonr) and
demonstrates that the nonresonant cross section is about the same
at this energy as the resonant production channel via $\Psi(3770)$
production and decay. Thus, in-medium modifications of the
$\Psi(3770)$ will be hard to detect experimentally via the $D
\bar{D}$ invariant mass spectrum (see below).

In analogy to Fig. 3 we show in Figs. 4 and 5 the calculated cross
section for the $\Psi(4040)$ at $T_{lab}$ = 6.8 GeV and
$\Psi(4160)$ at $T_{lab}$ = 7.7 GeV for the targets $^{12}C$,
$^{64}Cu$, $^{108}Ag$ and $^{207}Pb$ as a function of target mass
$A$. Due to the shorter life times of these resonances the
fraction of in-medium resonance decays increases up to 65 \% and
75\%, respectively, for a $Pb$-target. However, for the
rescattering cross sections involved (cf. Section 2) the life
times in the medium are shortened only by 8\% and 5\%,
respectively. This very moderate change of the spectral functions
due to in-medium interactions will be hard to see by dilepton
spectroscopy except for setups with excellent invariant mass
resolution.

\begin{figure}[h]
\centerline{\psfig{figure=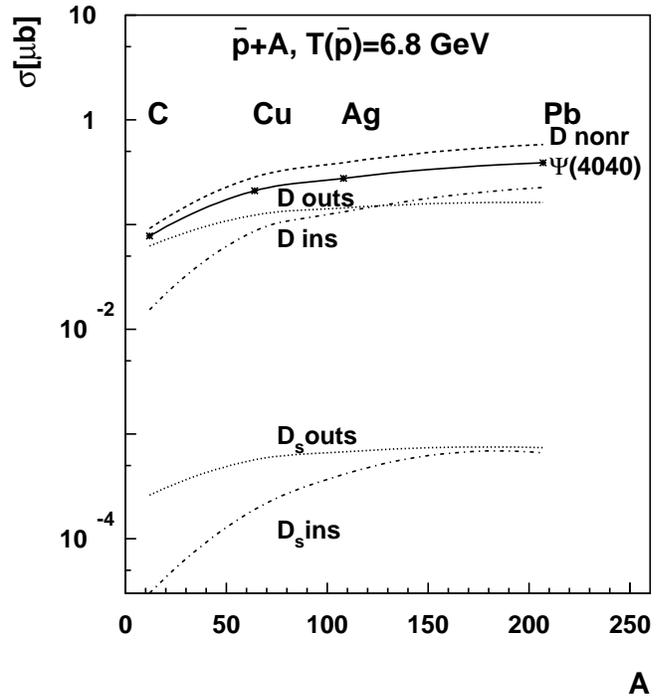,width=8.5cm}}
 \caption{The calculated cross sections for $\Psi(4040)$ at $T_{lab}$ = 6.8 GeV
 as a function of the target mass $A$. The dotted line (D outs) denotes the contribution
 from 'outside' resonance decays whereas the dash-dotted line (D ins) shows the
 resonance contribution from 'inside' decays. The solid line reflects the
 sum of 'outside' and 'inside'
 resonance decay contributions.
 The dashed line (D nonr) corresponds to the estimated nonresonant
 $D, \bar{D}$ cross section. The lower two lines give the 'outside' and 'inside' decay
 to $D_s,\bar{D}_s$, respectively.}
 \label{bild4}
\end{figure}

For the $\Psi(4040)$ the nonresonant $D, \bar{D}$ cross section
(upper dashed line; D nonr) becomes larger than the resonant cross
section (solid line) due to the higher invariant energy $\sqrt{s}$
involved in the elementary reactions at $T_{lab}$ = 6.8 GeV.
Furthermore, the resonant decays to $D_s, \bar{D}_s$ are
suppressed by almost 3 orders of magnitude in case of the
$\Psi(4040)$ which is not favorable for studying  $D_s, \bar{D}_s$
propagation in the nuclear medium. For the $\Psi(4160)$ an even
larger nonresonant contribution is expected which we do not
attempt to estimate explicitly since no educated guess for the
additional channels $\bar{D} D \pi +X$ is available. However, the
perspectives for $D_s, \bar{D}_s$ dynamics become more promising
due to larger branching ratios which are a consequence of the
higher invariant energy above the threshold and a more favorable
matrix element. Especially for the heaviest targets almost 50\% of
the $D_s, \bar{D}_s$ mesons from $\Psi(4160)$ decay appear inside
the nucleus such that rescattering effects can be explored. Such
rescattering phenomena can also be tested for nonresonant $D_s,
\bar{D}_s$ production channels; however, their cross section is
(at present) hard to estimate as mentioned before.

\begin{figure}[h]
\centerline{\psfig{figure=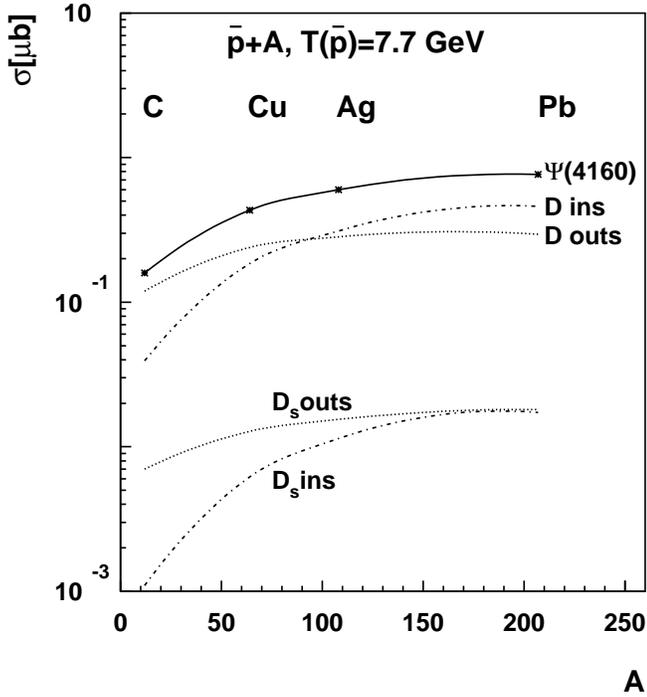,width=8.5cm}}
 \caption{The calculated cross sections for $\Psi(4160)$ at $T_{lab}$ = 7.7 GeV
 as a function of the target mass $A$. The dotted line (D outs) denotes the contribution
 from 'outside' resonance decays whereas the dash-dotted line (D ins) shows the
 resonance contribution from 'inside' decays. The solid line reflects the
 sum of 'outside' and 'inside' resonance decay contributions.
  The lower two lines give the 'outside' and 'inside' decay
 to $D_s,\bar{D}_s$, respectively.}
 \label{bild5}
\end{figure}

\section{Open charm propagation and rescattering}
The open charm mesons produced by resonance decay or nonresonant
channels may rescatter in the nuclear medium either elastically or
inelastically. In case of inelastic reaction channels there may be
charge exchange reactions such as ${D}^+ n \leftrightarrow {D}^0
p$ or $\bar{D}^- p \leftrightarrow \bar{D}^0 n$ or charm flavor
exchange reactions $D N \rightarrow \Lambda_c \pi$ or $\Sigma_c
\pi$ in analogy to the strangeness exchange reactions $\bar{K} N
\rightarrow \Lambda \pi $ or $\Sigma \pi$. Whereas charge exchange
reactions - induced by ($u,d$) quark exchange - do not change the
ratio ${D}/\bar{D}$ in nuclei, the charm exchange reaction leads
to a loss of $c$ quarks in the mesonic sector and thus to a
decrease of the ${D}/\bar{D}$ ratio with the target mass number
$A$.

In order to obtain some idea about the momenta of open charm
mesons in the laboratory we display in Fig. 6 the momentum
distribution of $D$ and $\bar{D}$ mesons - after full rescattering
- for $\bar{p} + Pb$ at $T_{lab}$ = 5.7 GeV. At this energy the
resonance contribution stems from the $\Psi(3770)$; its 'outside'
decay contribution is shown in Fig. 6 by the hatched area, which
extends from 2.2 GeV/c to about 4.2 GeV/c. The 'inside' resonance
contribution (denoted by $\bar{{\rm D}}$r and Dr, respectively) is
seen to be broadened significantly by elastic scattering and
charge exchange reactions showing also a sizeable net absorption
of $D$ mesons in case of the $Pb$ target of $\sim$ 30\%. On the
other hand, the $D, \bar{D}$ mesons from nonresonant production
channels signal larger in-medium rescattering effects which is
seen by comparison to the total $D$ (solid line) and $\bar{D}$
(dashed line) momentum spectra. The latter phenomenon is due to
the fact that rescattering of open charm mesons in case of
nonresonant production channels may proceed earlier than in case
of the $\Psi(3770) \rightarrow D\bar{D}$ channel since the
$\Psi(3770)$ on average is propagating for a couple of fm/c before
its decay.

\begin{figure}[h]
\centerline{\psfig{figure=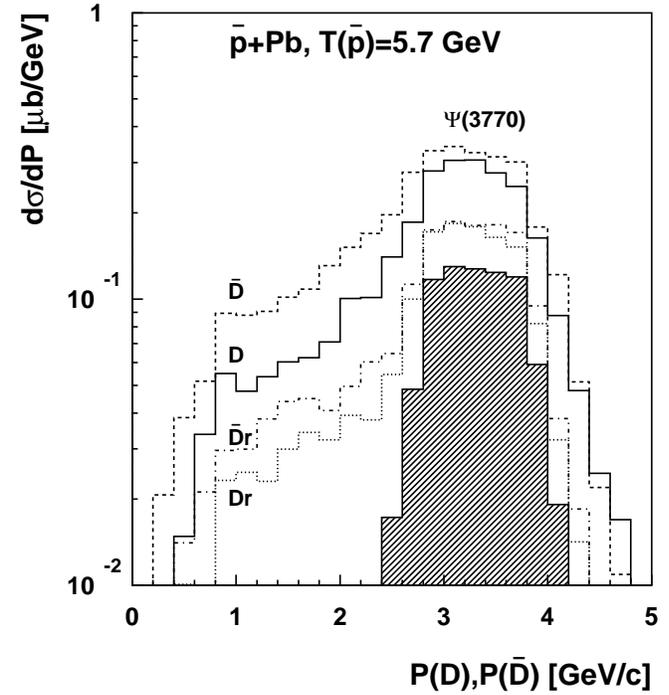,width=8.5cm}}
 \caption{The calculated momentum distributions of $D,\bar{D}$
 mesons for $\bar{p}+Pb$
reactions at $T_{lab}$ = 5.7 GeV. The hatched area reflects open
charm mesons from $\Psi(3770)$ decays in vacuum, the dotted (Dr)
and dash-dotted ($\bar{{\rm D}}$r) lines correspond to resonance
decays in the target nucleus, respectively. The solid and dashed
lines display the total spectra for $D$ and $\bar{D}$ mesons
including the nonresonant production channels (see text). In the
calculations  elastic and inelastic scatterings of the
$\Psi(3770)$ and $D, \bar{D}$ mesons have been taken into
account.}
 \label{bild6}
\end{figure}

In order to quantify the observable consequences from $D, \bar{D}$
rescattering and absorption we have made a cut on the laboratory
momenta $P_D, P_{\bar{D}} \leq 2.2$ GeV/c to exclude the vacuum
decays of the $\Psi(3770)$ and to reduce the fraction of
unscattered $D, \bar{D}$ events. The calculated cross sections for
resonant (l.h.s.) and nonresonant (r.h.s.) production channels are
shown in Fig. 7 (upper part) for $D$ and $\bar{D}$ mesons,
respectively, as a function of the target mass $A$ at $T_{lab}$ =
5.7 GeV. Even for the low momentum cut the cross sections are in
the order of 50 -- 100 $nb$ for heavy targets, which should be
feasible at the HESR with a $\bar{p}$ luminosity of $\sim$
2$\times$ 10$^{32}$. The resulting ratio ${D}$ to $\bar{D}$ mesons
is displayed in the lower part of Fig. 7 for resonant (l.h.s.) and
nonresonant (r.h.s.) productions separately. The constant line
(=1) results trivially when including charge exchange reactions,
however, discarding $c$-quark exchange reactions. As noted above,
the ${D}$-meson absorption is larger in the nonresonant case and
may reach 40\% for the $Pb$-target within the cross sections
specified in Sec. 2. Vice versa, experimental data on this ratio
(employing the same low momentum cut) as a function of mass $A$
will allow to set stringent bounds on the size of the $c$-quark
exchange cross section in the nuclear medium.

\begin{figure}[h]
\centerline{\psfig{figure=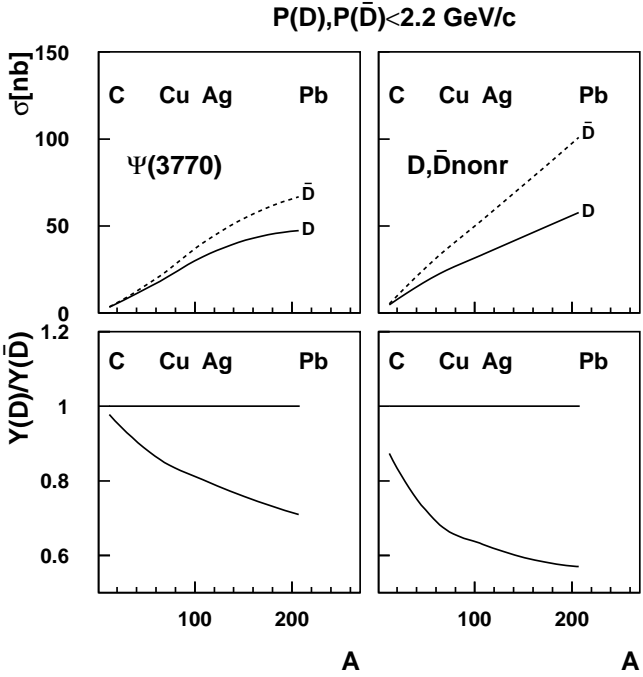,width=8.5cm}}
 \caption{Upper part: The calculated cross sections for $D$ and $\bar{D}$ mesons
 for $\bar{p}+A$ reactions at $T_{lab}$=5.7 GeV including a low
 momentum cut $P_D$, $P_{\bar{D}} \leq$
2.2 GeV/c. The l.h.s. shows the contribution from in-medium
resonance decays whereas the r.h.s. corresponds to nonresonant
production channels. Lower part: The ratio $D/\bar{D}$ versus the
target mass $A$ for resonant (l.h.s.) and nonresonant (r.h.s.)
production channels. The constant line (=1) results when
neglecting $c$-quark exchange channels.}
 \label{bild7}
\end{figure}

Without explicit representation we note that the perspectives of
measuring $D, \bar{D}$ rescattering in heavy nuclei from the
resonance decay of the $\Psi(4040)$ are similar to the previous
case discussed above and also close to the results for the
$\Psi(4160)$ (see below). Since the chances to measure
additionally the re\-scattering effects of $D^+_s, D^-_s$ mesons
are very low in view of the small partial decay width for the
$\Psi(4040)$ (cf. Fig. 1) we directly continue with a discussion
of the $\Psi(4160)$ formation and decay at $T_{lab}$ = 7.7 GeV.

\begin{figure}[h]
\centerline{\psfig{figure=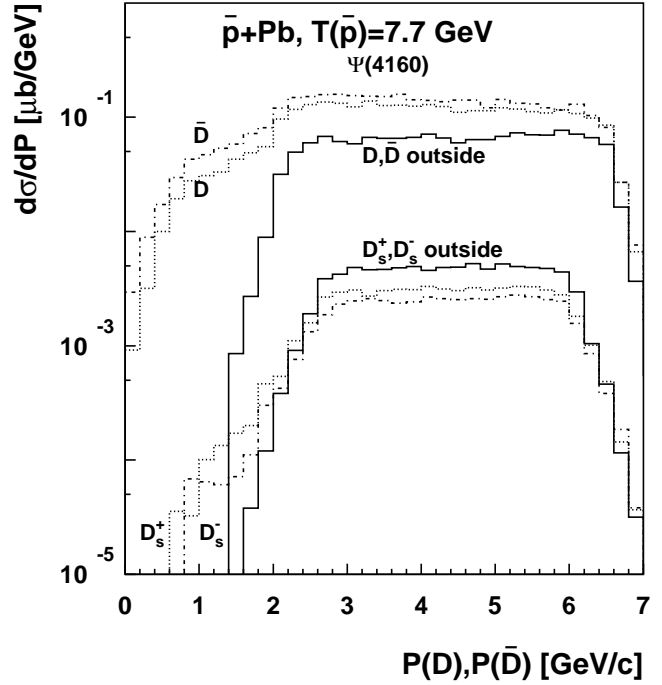,width=8.5cm}}
 \caption{The calculated momentum distributions of $D,\bar{D}$
 (upper lines) and $D_s^+,D_s^-$ (lower lines)
 mesons for $\bar{p}+Pb$
reactions at $T_{lab}$ = 7.7 GeV. The solid lines ($D, \bar{D}$
outside, $D^+_s, {D}^-_s$ outside) reflect open charm mesons from
$\Psi(4160)$ decays in vacuum, the dotted and dash-dotted
($\bar{D}$) lines correspond to resonance decays in the target
nucleus, respectively.  In the calculations elastic and inelastic
scatterings of the $\Psi(4160)$, $D, \bar{D}$ and $D_s^+,D_s^-$
mesons have been taken into account.}
 \label{bild8}
\end{figure}

In this case (for $\bar{p} + Pb$) the momentum distribution of the
$D, \bar{D}, D^+_s, D^-_s$ mesons becomes very broad and extends
from $\sim 0$ to $\sim$ 7 GeV/c (cf. Fig. 8) where the 'outside'
$\Psi(4160)$ decays (solid lines; $D,\bar{D}$ outside,
$D^+_s,D^-_s$ outside) give open charm mesons essentially above
$\sim$ 2 GeV/c. Thus rescattering events of the $D$-mesons are
dominantly found for momenta less than 2 GeV/c which,
unfortunately, are only a low fraction of the total open charm
yield. Furthermore, the number of $D^+_s, D^-_s$ events in this
low momentum range is down by more than 2 orders of magnitude
relative to the $D, \bar{D}$ mesons. However, apart from the
resonant production of charm meson pairs there will be an
additional contribution from nonresonant $D, \bar{D}, D^+_s,
D^-_s$ mesons which will enhance the cross section also for low
momenta, respectively. It is hard to provide reliable estimates
for the nonresonant contribution, but in view of Fig. 4 for
$T_{lab}$ = 6.8 GeV this contribution might be higher by a factor
2--4 than the resonance contribution.

As shown in Fig. 9 the cross sections for $D, \bar{D}$ mesons at
$T_{lab}$ = 7.7 GeV are above 50 $nb$ for heavy targets (l.h.s,
upper part) -- even when including a low momentum cut of 2 GeV/c
-- showing a sizeable difference due to the absorption of $D, D^*$
mesons in the nucleus. The perspectives for $D^+_s, D^-_s$ mesons
are less promising (r.h.s, upper part). Here only cross sections
of $\sim$ 0.1 $nb$ should be expected for a $Pb$ target which
might be enhanced due to nonresonant channels by a factor 2--4 (as
estimated above). However, the relative absorption of $D,D^*$
mesons -- as measured by the ratio to $\bar{D}, \bar{D}^*$ mesons
-- increases with target mass by up to 35\% for $A$=207 (l.h.s.,
lower part) for the cross sections specified in Sect. 2. This
decrease in the $D/\bar{D}$ ratio is roughly the same for the
$\Psi(4160)$ decays as for the $\Psi(3770)$ decays (Fig. 7)
inspite of the shorter lifetime of the $\Psi(4160)$. This is
partly due to the different cut in momentum and the fact, that the
average momentum of the produced $\Psi(4160)$ in the laboratory is
higher. The ratio of $D^-_s/D^+_s$ versus target mass $A$ (r.h.s.;
lower part) also shows a decrease due to the different cross
sections for $c$ and $s$ quark exchanges with the nucleons of the
target. The vertical lines in Fig. 9 show the statistical error
bars in this ratio for the low momentum cut of 2 GeV/c that
reflect the low number of $D^-_s, D^+_s$ mesons in this momentum
range as obtained from the Monte Carlo simulations. In principle,
provided that sufficient statistics are obtained experimentally,
it is possible to determine the relative strength of $(u,d)$, $s$
and $c$ quark exchanges with nucleons from the mass dependence of
the further ratios $D/D^+_s$ and $D/D^-_s$. We only point out this
possibility without presenting explicit figures for these ratios
since the statistics for $D_s^{\pm}$ is very small and the cross
sections employed are rather uncertain so far.

\begin{figure}[h]
\centerline{\psfig{figure=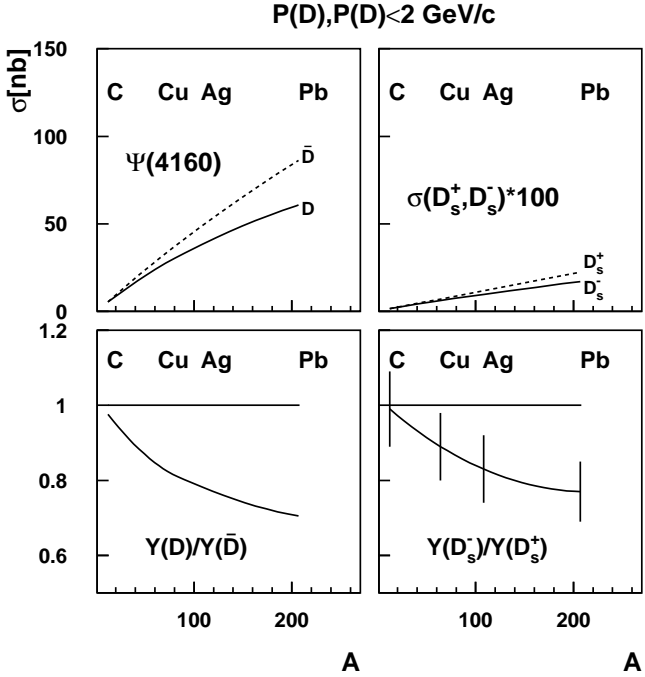,width=8.5cm}}
 \caption{Upper part: The calculated cross sections for $D$ and $\bar{D}$ mesons
 for $\bar{p}+A$ reactions at $T_{lab}$=7.7 GeV including a low
 momentum cut $P_D$ and $P_{\bar{D}} \leq$
2.0 GeV/c. The l.h.s. shows the contribution from in-medium
resonance decays for $D, \bar{D}$ whereas the r.h.s. corresponds
to $D^+_s, {D}^-_s$ resonant production (multiplied by a factor
100). Lower part: The ratio ${D}/\bar{D}$ versus the target mass
$A$ (l.h.s.) and the ratio ${D}^-_s/{D}^+_s$ for resonant
production. The constant line (=1) results when neglecting $s$ and
$c$ quark exchange channels.}
 \label{bild9}
\end{figure}

\section{Summary} In this study we have explored the perspectives of
measuring the dynamics of hidden charm vector mesons and open
charm mesons in nuclei in antiproton induced reactions on nuclei.
Such experimental studies will be possible at the high-energy
antiproton storage ring HESR proposed for the future GSI facility
\cite{GSIfut}. The Multiple Scattering Monte Carlo (MSMC)
calculations performed have been based on the resonance production
concept with resonance properties from the PDG \cite{PDG} or
related estimates from high energy open charm branching ratios. We
note, that these resonance properties are in line with
$\Psi(3770), \Psi(4040)$ and $\Psi(4160)$ production from $e^+
e^-$ annihilation \cite{Bai}, however, the detailed branching
ratios for the $\Psi(4040)$ and $\Psi(4160)$ (cf. Fig. 1) are
still quite uncertain. Thus our estimated uncertainty in the cross
sections is in the order of a factor 2--3, which should be kept in
mind when planning experiments along this line. We stress that the
total branching ratios to $\bar{p}p$ should be measured first with
high accuracy. Deviations from our calculations for hidden charm
vector mesons in $\bar{p}p$ reactions than directly carry over to
the cross sections on nuclei (Fig. 2). Nevertheless, even for a
light nucleus such as $^{12}C$ cross sections from 70 -- 150 $nb$
should be expected at the proper resonance energies (Fig. 2). Our
calculations suggest that the total yield of these vector mesons
scales as $\sim A^{0.55}$ with target mass $A$ (Fig. 3). For a
heavy nucleus like $Pb$ the fraction of 'inside' decays to
dileptons or $D, \bar{D}$ reaches $\sim$ 50\% for the
$\Psi(3770)$, $\sim$ 65 \% for the $\Psi(4040)$ and $\sim$ 75\%
for the $\Psi(3770)$, which is sufficient to address the question
of in-medium properties of these resonances. However, for the
rescattering cross sections adopted in Sect. 2, the in-medium
modifications of their spectral functions is rather moderate, i.e.
15\%, 8\% and 5\% change in width, respectively, such that a mass
resolution for dilepton pairs in the order of a few MeV will be
necessary. This will be also the case for in-medium mass shifts of
these vector mesons which are estimated to be in the order of a
few MeV, only \cite{Klingl}.

We have, furthermore, explored the perspectives of measuring the
interactions of open charm mesons in nuclei that are produced by
resonant vector meson decays or nonresonant channels,
respectively. For the cross sections specified in (12) the effects
of $D,\bar{D}$ rescattering and $D$-meson absorption in heavy
nuclei can clearly be identified when imposing a low momentum cut
in the laboratory of 2--2.2 GeV/c at $T_{lab}$ = 5.7 or 7.7 GeV,
respectively. This low momentum cut excludes events with $D
\bar{D}$ decays in vacuum or $D, \bar{D}$ mesons that did not
rescatter in the target nucleus. The cross sections for such type
of events with $D$ or $\bar{D}$ mesons of low momenta are in the
order of 50 $nb$ for a heavy target like $Pb$. Such cross sections
are high enough to perform successful experiments with an
antiproton luminosity of 2$\times$ 10$^{32}$ as envisaged for the
HESR \cite{GSIfut}. Experimentally, the ratio of $D/\bar{D}$ (at
low momenta) as a function of target mass $A$ allows to set
stringent constraints on the $D$-meson absorption -- or $c$-quark
transfer -- cross section at nuclear density $\rho_0$.
Furthermore, at $T_{lab} \approx$ 7.7 GeV the production cross
section of $D^+_s, D^-_s$ mesons should become appreciable and be
in the order of 20 $nb$ for a $Pb$ target when including only the
$\Psi(4160)$ decays. Approximately 50\% of these ($c\bar{s}$) and
($\bar{c} s$) mesons should rescatter in a target of mass $A
\approx$ 200 such that $c$ and $s$-quark exchange reactions with
nucleons can be measured by studying the mass dependence of
$D^-_s/D^+_s$ as well as $D/D^-_s$ and $D/D^+_s$ at low momenta.
However, to  exclude events from $\Psi(4160)$ vacuum decays as
well as events with unscattered open charm mesons a severe cut on
low momenta ($\leq 2$ GeV) should be performed. In this momentum
range the cross section for $D^-_s, D^+_s$ drops to $\sim$ 0.1
$nb$ even for a $Pb$ target. It will thus be very hard to measure
the various particle ratios as a function of mass $A$ with good
statistics.

\section*{Acknowledgements} We are grateful to A. B. Kaidalov
and A. Sibirtsev for helpful discussions and valuable suggestions.
Furthermore, we like to thank E. L. Bratkovskaya and U. Wiedner
for critical comments and a careful reading of the manuscript.
This work was supported by DFG under grant No. 436 RUS 113/600 and
RFFI.

\end{document}